\documentclass[aps,prl,twocolumn,groupedaddress]{revtex4}
\pdfoutput=1
\usepackage{graphicx}
\usepackage{bm}
\usepackage{amsmath}
\def\H{\mathrm{H}}
\def\s{{\bf s}}

\def\up{\uparrow}
\def\down{\downarrow}
\begin{document}


\title{Molecular spintronics using noncollinear magnetic molecules}


\author{Alessandro Soncini}
\email{Alessandro.Soncini@chem.kuleuven.be}
\author{Liviu F. Chibotaru}
\affiliation{Institute for Nanoscale Physics and Chemistry (INPAC) and Division of Quantum and Physical Chemistry, Katholieke Universiteit Leuven, Celestijnenlaan 200F, B-3001 Heverlee, Belgium}

\date{\today}

\begin{abstract}

We investigate the spin transport through strongly anisotropic noncollinear magnetic molecules and find that 
the {\em noncollinear magnetization} acts as a spin-switching device for the current. 
Moreover, spin currents are shown to offer a viable route to selectively prepare
the molecular device in one of two degenerate noncollinear magnetic states.
Spin-currents can be also used to create a non-zero density of toroidal magnetization
in a recently characterized Dy$_3$ noncollinear magnet.
\end{abstract}
\pacs{86.65.+h,72.25.-b,75.30.Gw,75.50.Xx}

\maketitle

One of the most ambitious directions in the quest for the ultimate miniaturization of 
electronic devices is represented by molecular spintronics~\cite{sanvito,bogani}.
Molecular nanomagnets are particularly promising for nanospintronics, especially
in relation to the quest for magnetic molecular qubits~\cite{TimcoNature}, since
transport experiments~\cite{experiment1,experiment2,experiment3} have shown
a strong interplay between the current and the magnetic states of the molecules.
To date, all theoretical investigations on molecular spintronics have addressed 
systems whose magnetism is only weakly anisotropic, thus exploring systems whose magnetization 
aligns along a single anisotropy axis (collinear magnetism)~\cite{elste,romeike,gonzales,misiorny}.  
The noncollinear regime of molecular magnetism, arising when the on-site magnetic anisotropy of single metal
ions is one of the dominant energy scales, has only been explored very recently~\cite{prbrapid}. 
In the noncollinear regime magnetic molecules can be prepared in degenerate states 
characterized by non-dipolar magnetic moments, such as the recently predicted~\cite{prbrapid} and 
found~\cite{dy3,prl} toroidal (or anapole) moment in molecular wheels.

There are two main arguments that make spintronics of noncollinear magnets 
of great interest. The first follows from studies 
on spin-transport through mesoscopic rings with noncollinear internal magnetic fields,
which have been predicted to produce spin-switching effects~\cite{frustaglia}.
The size of noncollinear molecular scatterers is expected to be more favorable to
overcome dephasing, and lead to the observation of coherence and spin-switching effects.
The second argument is related to the use of noncollinear states to implement molecular qubits.  
On the one hand, molecular spin-qubits can easily be addressed via a magnetic field~\cite{troiani}, 
although intermolecular dipolar interactions lead to short dephasing times~\cite{morello}. On the other hand, 
intermolecular interactions between non-dipolar states are weak~\cite{prbrapid}, thus decoherence times 
longer, although these states cannot be addressed via uniform fields. Spintronics might offer 
a promising strategy to address noncollinear protected molecular qubits.

In this Letter we investigate spin transport through molecular noncollinear magnetic states, 
and provide evidence that these systems do offer strategies to 
(i) implement quantum-interference molecular devices capable of reversing the 
polarization of an injected spin-current, and to (ii) selectively populate specific 
noncollinear magnetic states. 
The most relevant transport regime has been shown to be the Coulomb blockade (CB)~\cite{experiment3}. 
\begin{figure}
\begin{center}
 \scalebox{0.18}{%
 \includegraphics*{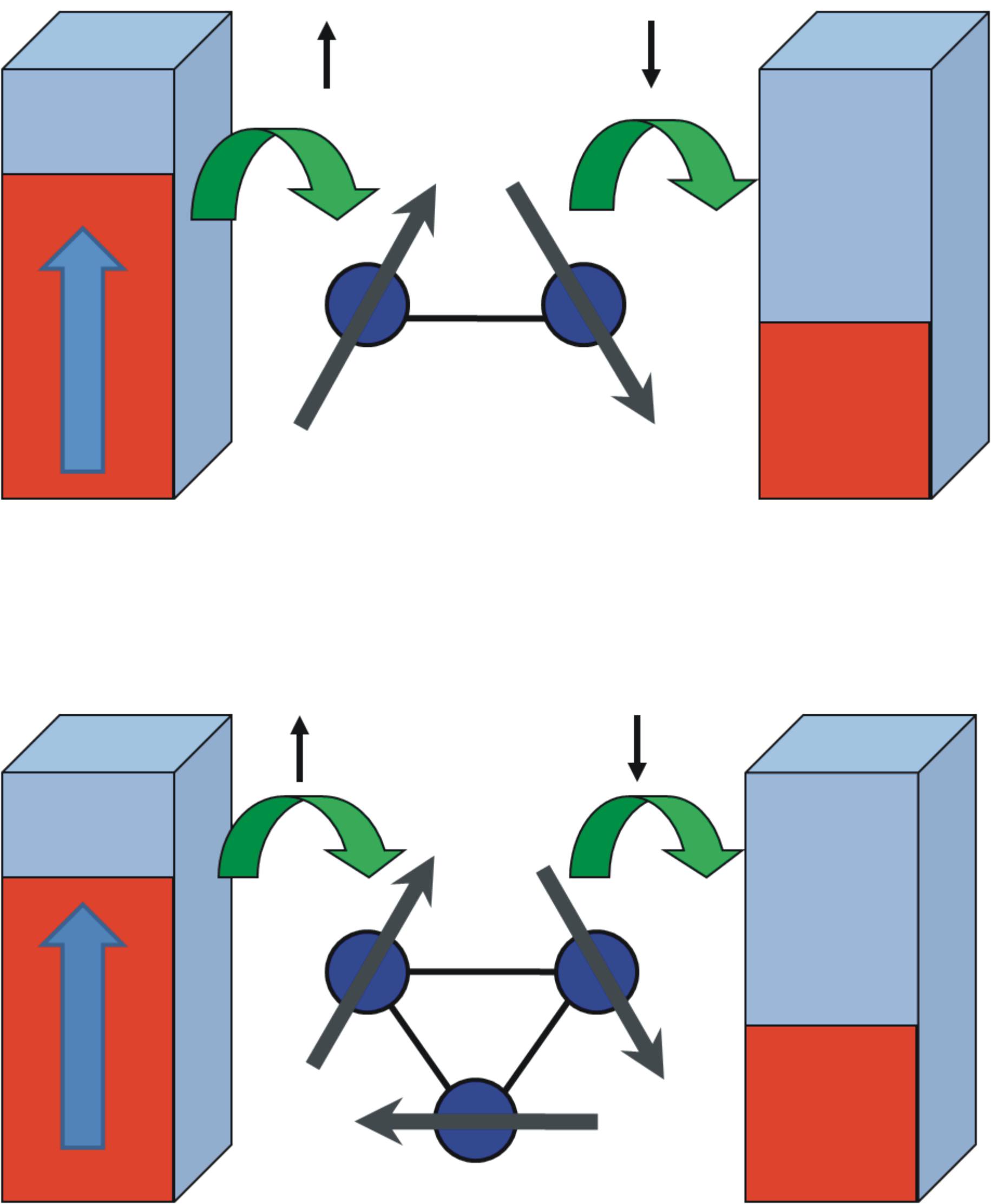}}
\end{center}
\caption{(color online) The two molecular spintronics setups considered in this work. Top, a noncollinear 
antiferromagnetic dimer with on-site spin $s=3/2$ and coplanar on-site ZFS easy-axes, tilted with respect to the 
perpendicular to the inter-metal distance by $\pm 30^\circ$. Bottom, a 3-center antiferromagnetic molecular wheel with
on-site spin $s=3/2$  and coplanar on-site ZFS easy-axes, arranged tangentially to the wheel's circumference (cf Ref. \cite{prbrapid,dy3}).
\label{fig1}}
\end{figure}
The lowest lying states of a nanomagnet with $n$ unpaired electrons well localized on $N$ metal 
centers with local spin $s$ is well described by the Hamiltonian:
\begin{eqnarray}
\H_{n} & = & -J \sum_{\left\langle ij \right\rangle} {\tilde \s}_i \cdot {\tilde \s}_{j} 
+  D \sum_{i} {\tilde s}_{z,i}^{2}
\label{hex}
\end{eqnarray}
consisting of the isotropic Heisenberg exchange coupling between nearest neighbors with strength $J$, 
and easy-axis zero-field splitting (ZFS) on-site (strength $D$ with $D<0$). Note that the spin
operator ${\tilde {\bm s}}_i$ has $z$-component ${\tilde s}_{i,z}$ parallel to the local ZFS axis. 
Whereas previous investigations~\cite{misiorny} only considered the collinear weak-anisotropy regime, 
here we introduce two key-ingredients for {\em noncollinearity}~\cite{prbrapid}: (i) $|D|>>|J|$
(ii) on-site easy-axes {\em not parallel} to each other. 
In this work we will explore spin-transport for a dimer (Fig. 1, top) 
and a three-centers molecular wheel (Fig. 1, bottom).

When connected to source, drain and gate electrodes, under bias voltage $V_B$ and gate voltage $V_G$, 
the molecule will become charged. The migrating electron will be consecutively accommodated 
at different metal sites~\cite{jacs}, described here by a set of $N$ atomic orbitals localized on the metal centers. 
The molecular Hamiltonian for a charged state with an excess of $Q$ electrons with respect 
to the isolated nanomagnet is given by:
\begin{eqnarray}
\H_{n+Q} & = & \H_{n} + (\epsilon -eQV_G)\sum_p^{N}\sum_{\sigma}^{\up\down} n_{p\sigma}\nonumber\\
         & + & t\sum_{\left\langle p q \right\rangle}\sum_{\sigma}^{\up\down} c_{p\sigma}^{\dagger}c_{q\sigma}
+U\sum_p^{N} n_{p\up}n_{p\down}\nonumber\\
&+&J_{\mathrm{H}}\sum_{p}^{N}\sum_{\alpha}^{\up\down}\sum_{\beta}^{\up\down} 
{\tilde {\bm s}}_p \cdot {\bm \sigma}_{p,\alpha\beta}c_{p\alpha}^{\dagger}c_{p\beta},
\label{Hcharge}
\end{eqnarray}
where $\epsilon$ is the energy of the localized orbitals, $c_{p\sigma}^\dagger$ are creation 
operators for the on-site spin-orbitals, $n_{p\sigma}=c_{p\sigma}^{\dagger}c_{p\sigma}$, 
$t$ is a hopping parameter between centers, $U$ is the Coulomb repulsion between two electrons on the same 
center, ${\bm \sigma}_{p}$ are Pauli matrices associated to an electronic spin injected on site $p$, 
and $J_{H}$ is the Hund's rule coupling between the spin of the excess electron on site $p$, and the spin moment 
${\tilde {\bm s}}_p$ on that center ($J_{H}<0$). 
Here we confine ourselves with the region around the first CB diamond, where only singly-charged states are relevant 
together with the neutral ones. This is formally achieved by setting $U \rightarrow\infty$.
Finally, a tunneling Hamiltonian $\H_{\mathrm{mix}}$ between electrodes and device is introduced in the 
usual manner~\cite{elste,romeike,gonzales,misiorny}, with tunneling amplitudes estimated to be at 
most 0.3 cm$^{-1}$~\cite{gonzales,experiment3}.
Given the weak molecule-lead coupling, the transition rates $W$ between molecule and contacts
are calculated with the Fermi golden rule using $\H_{\mathrm{mix}}$, assuming a Fermi-Dirac
distribution in the two leads, kept at different chemical potential $\mu_L - \mu_R =  e V_B$.
Next, using the rates $W$, a master equation for the non-equilibrium populations of 
charged and neutral states of the device is set up and solved
in the steady-state regime.  The resulting populations are used to compute the input 
($I_L^{\up}-I_L^{\down}$) and the output ($I_R^{\up}-I_R^{\down}$) spin currents.  
\begin{figure}
\begin{center}
 \scalebox{0.48}{%
 \includegraphics*{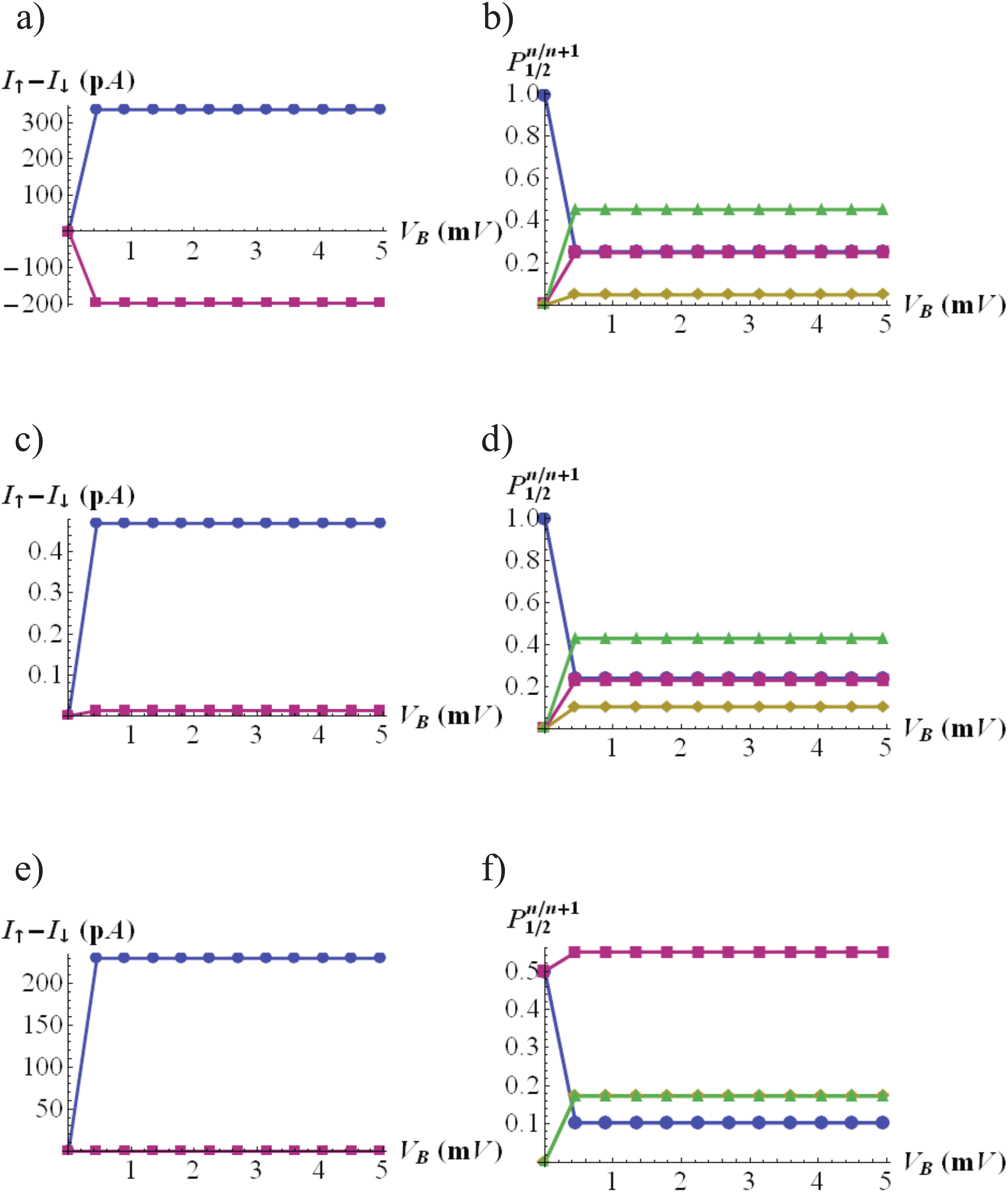}}
\end{center}
\caption{(color online) Input (blue bullet datapoints) and output (purple square datapoints) 
spin-current vs. voltage curves (left column), and, (right column) non-equilibrium populations of 
neutral (blue bullet for in-phase and  purple square for out-of-phase superposition of
$\left|+-\right\rangle$ and $\left|-+\right\rangle$) and 
charged states (green upward-pointing triangles for $\Psi_{0,1}^{n+1}$, yellow diamonds for $\Psi_{0,2}^{n+1}$)
 for the noncollinear magnetic dimer, with the following type of
exchange interaction and values of the transfer parameters ($D = -250$ cm$^{-1}$): 
a) and b) Heisenberg, $t=0.5 D$,
c) and d) Heisenberg, $t=4.0 D$,
e) and f) noncollinear Ising, $t=1.0 D$.
\label{fig1}}
\end{figure}

Let us first consider the dimer molecule with local spin $s=3/2$ ({\it e.g.}, a Co(II) dimer). 
We assume co-planar local ZFS axes, forming angles $\theta=\pm 30^{\circ}$  with the perpendicular 
($z$-direction) to the Co-Co bond, (Fig. 1, top).  We choose here $D= 5 J$, with antiferromagnetic 
isotropic exchange coupling $J=-50$ cm$^{-1}$.  The dominant energy scale in (\ref{Hcharge}) is 
the on-site Hund coupling exchange $J_H$, chosen here as $J_H = 4 D \approx 0.1 eV$.  Moreover,
the source contact is assumed to be ferromagnetic, the drain non-magnetic. The spin-polarization 
axis for the ferromagnetic source is coplanar to the ZFS axes, and parallel to $z$ (Fig. 1). 
The Heisenberg states of lower energy of the Co-dimer can be described in terms of almost 
pure noncollinear Ising states $\left|m_1 m_2\right\rangle$, where $m_i$ is 
the projection of the local spin moment $s$ along the tilted easy axis \cite{prbrapid}.
This is verified by decomposing the Heisenberg wavefunction into the noncollinear Ising basis.
For the present choice of parameters, denoting ``$+$'' and ``$-$'' the on-site 
spin-states  $\left| \pm 3/2\right\rangle$, the ground state is quasi-degenerate and 
corresponds to the in-phase and out-of-phase superposition of the noncollinear Neel 
states $\left|+-\right\rangle$, and  $\left|-+\right\rangle$.  The tunneling gap 
is about  $\Delta\approx 0.38$ cm$^{-1}$.

Next, we find the eigenstates of~(\ref{Hcharge}). 
We explore here two limiting situations: a weak-transfer limit with $t = 0.5 D$, and 
a strong-transfer limit with $t=4 D = J_H$.
In the weak-transfer limit, we expect the additional electron spin to 
follow ``adiabatically'' the noncollinear magnetic texture of the molecular device, 
so that the ground state of the charged system will be doubly degenerate, and
dominated by either the $\left|+-\right\rangle$ or the  $\left|-+\right\rangle$ 
component of the Neel doublet, carrying an additional electron on either center,
with the spin parallel to the local magnetic moment via Hund-rule coupling.
These expectations are confirmed by full diagonalization, leading to the following 
ground state for the $n+1$-magnet:
\begin{eqnarray}
\Psi_{0,1}^{n+1}  & \approx & C_{\uparrow,0} \left| +-\right\rangle \left| \uparrow 0 \right\rangle
+C_{0,\downarrow}  \left| +-\right\rangle \left| 0 \downarrow\right\rangle \nonumber \\
\Psi_{0,2}^{n+1}  & \approx & C_{\down,0} \left| -+\right\rangle \left| \down 0 \right\rangle
+C_{0,\up}  \left| -+\right\rangle \left| 0 \up\right\rangle, 
\label{adiabatic_transport}
\end{eqnarray}
where the kets $\left| \uparrow 0 \right\rangle$ denote the determinant of two spin-orbitals
centered on the two metals, the first with spin $\up$, the second empty. 
We find $\left|C_{\uparrow,0}\right|^2= \left|C_{\down,0}\right|^2 \approx 0.45$, and
$\left|C_{0,\downarrow}\right|^2 = \left|C_{0,\up}\right|^2\approx 0.42$.
On the other hand, in the strong-transfer limit, inter-center hopping processes have the 
same rate as Hund-coupling spin-polarization processes, so that transport 
is not expected to be adiabatic. This is confirmed 
by full diagonalization, where the ground state is dominated by noncollinear
Ising states favoring spin-preserving hopping processes, such 
as $\left| ++\right\rangle \left| \uparrow 0 \right\rangle$ and 
$\left| ++\right\rangle \left| 0 \uparrow \right\rangle$. Thus, in this regime
the overlap with the neutral noncollinear Neel states is very small.
The voltage $V_G$ is taken large enough to bring in 
resonance the ground states of the neutral and charged systems, separated by
about $1eV$.
Without loss of generality, we set the equilibrium chemical potential
lying in between the ground and first excited state of the neutral molecule,
and the temperature to $T=0.1K$.

In Fig. 2a (weak-transfer) and 2c (strong-transfer) we report the spin current-voltage 
diagrams obtained for the two limits of the hopping parameter. Since the 
source is fully spin-polarized, the input spin current (blue bullets)
always corresponds to the total charge-current.  Interestingly, in the weak-transfer 
limit the output spin-current (purple square datapoints) 
has a negative sign: {\it the spin polarization of the input current is
reversed in the output non-magnetic electrode, by the noncollinear magnetic texture}. 
On the other hand, in the strong-transfer regime this {\it spin-switching} effect is 
not observed.  These results are easily interpreted analyzing the ground state 
wavefunctions for the neutral and charged states. 
In the weak-transfer limit, the charged ground state (\ref{adiabatic_transport}) is a 
coherent state describing the adiabatic hopping  of an injected electron-spin between
the two metals, in which process the additional electron always aligns its spin parallel to
the magnetic polarization of the local metal ion.  Thus we define this limit as the 
{\it adiabatic-transport limit}, in analogy with the findings reported in ref.~\cite{frustaglia}
Although the CB-regime is non-coherent, the transition rates entering the master equation 
are determined by the overlap amplitudes between the tunneling combinations of the
ground noncollinear Neel doublet, and the charged ground-state doublet (\ref{adiabatic_transport}).
Due to the full $\up$-spin polarization of the source,
the injected electron on the first metal center creates an excess 
of non-equilibrium population in the state $\Psi_{0,1}^{n+1}$ (see Fig. 2b, green 
triangular datapoints), which can host an electron with spin-up on the first metal.
The electron is then coherently transported through $\Psi_{0,1}^{n+1}$ on the second metal 
center, where, as described by the  $\left| 0 \down\right\rangle$ component of $\Psi_{0,1}^{n+1}$,
its spin-polarization is reversed. Output tunneling events from the second metal center
into the drain will thus occur more frequently with opposite spin-polarization.
Since the additional spin-polarized electron collapses the tunneling wavefunction 
into one of the two Neel states, at non-zero bias voltage we note that the ground state
and first excited tunneling states of the neutral system become equally populated
(see Fig. 2b and 2d).

We note that the coupling between $\left| +-\right\rangle \left| \uparrow 0 \right\rangle$ and 
$\left| +-\right\rangle \left| 0 \down \right\rangle$  in (\ref{adiabatic_transport}), 
which determines the spin-switching transport, is triggered by the Hund-Hamiltonian.
Importantly, if the angle $\theta$ is set to zero, {\it i.e.} within the collinear regime,
the Hund-mechanisms leading to the superposition (\ref{adiabatic_transport}) are not active.  
Hence, {\it noncollinearity is found to be a crucial ingredient for the realization 
of the spin-switching effect}.  In the non-adiabatic regime, the spin-switch effect is quenched (Fig. 2c)
due to the negligible presence of spin-switch coherences in the charged ground state.
In this limit the large hopping integral favors spin-preserving hopping processes.
The current magnitude is also significantly smaller due to the small 
overlap between charged and neutral states (see Fig. 2c).

It is interesting to investigate the case of exact degeneracy between 
$\left| +-\right\rangle $ and  $\left| -+\right\rangle $, implied 
by the noncollinear Ising exchange Hamiltonian:
\begin{eqnarray}
\H_{n} & = & -J_I \sum_{\left\langle ij \right\rangle} {\tilde  s}_{i,z} {\tilde s}_{j,z}
+  D \sum_{i} {\tilde s}_{z,i}^{2}.
\label{hex}
\end{eqnarray}
The ground state of the charged system for $t= D$ is still doubly degenerate and given
by (\ref{adiabatic_transport}), with 
$\left|C_{\uparrow,0}\right|^2 = \left|C_{0,\downarrow}\right|^2 = \left|C_{\downarrow,0}\right|^2 
= \left|C_{0,\uparrow}\right|^2 = 0.41$.
The full $\up$-spin polarization of the source electrode will favor population-transfer 
processes mainly between the $\left| +-\right\rangle$ Neel state and the {\it spin-switch} 
excited state $\Psi_{0,1}^{n+1}$, via the Hund-coupling mechanism. 
However, now only half of the input $\up$-current is converted to $\down$-current,
so that the current is non-polarized in the drain (see Fig 2e).  This can be rationalized 
in terms of additional population-transfer between $\left| -+ \right\rangle $ and 
$\Psi_{0,2}^{n+1}$, since the latter contains $7\%$ of the Hund-unstable component
$\left| -+\right\rangle  \left| \up 0\right\rangle$.
\begin{figure}
\begin{center}
 \scalebox{0.48}{%
 \includegraphics*{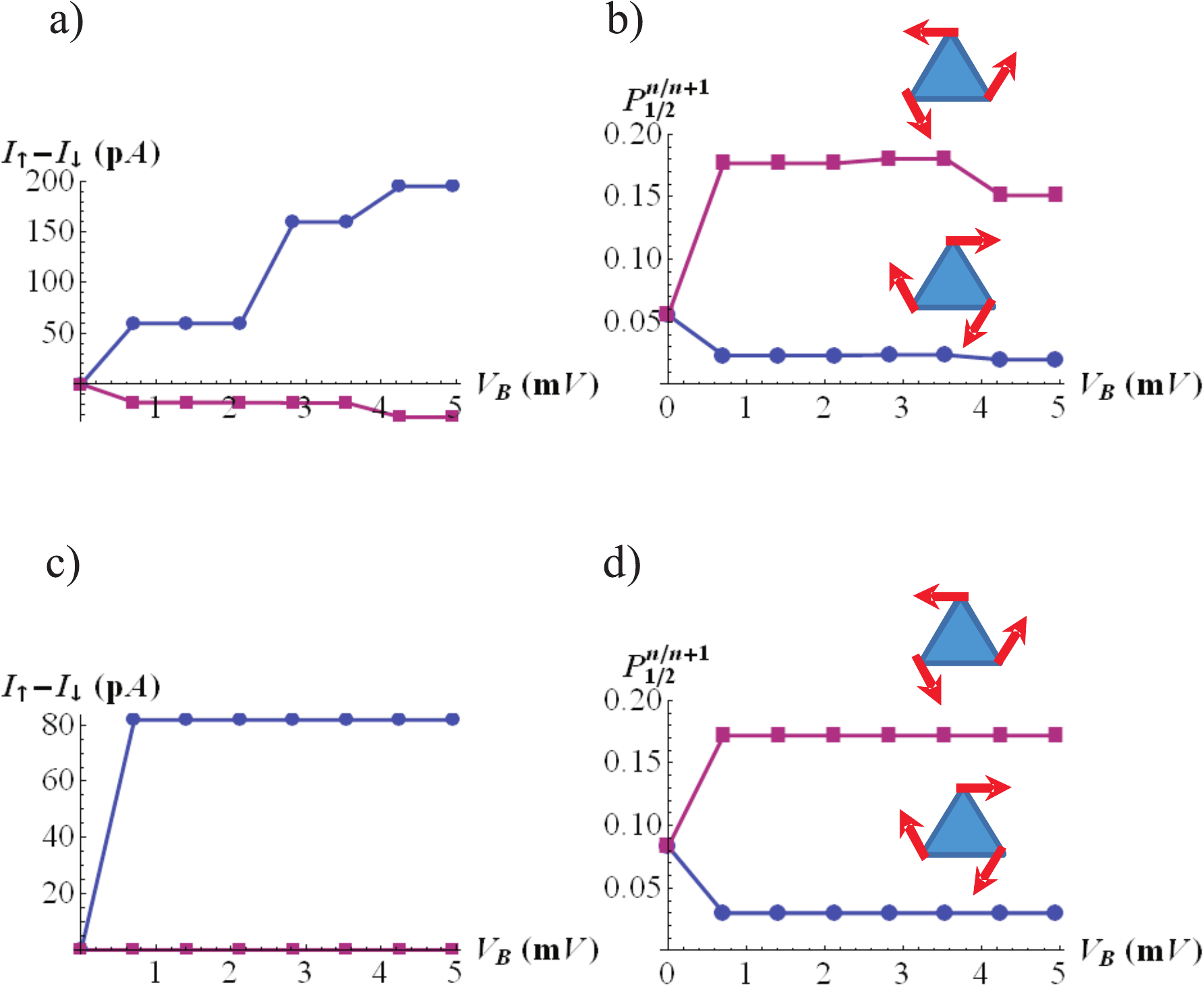}}
\end{center}
\caption{(color online) Input (blue bullet datapoints) and output (purple square datapoints) 
spin-current vs. voltage curves (left column), and, (right column) non-equilibrium populations 
of neutral (blue bullet for $\left|---\right\rangle$ and purple square for $\left|+++\right\rangle$) 
for the 3-center noncollinear magnetic wheel, with the following values of the 
transfer parameters: a) and b) $t=0.05 D$, c) and d) $t=1.0 D$ ($D = -200$ cm$^{-1}$).
\label{fig1}}
\end{figure}
However, the dominant population-transfer process 
remains  $\left| +-\right\rangle \rightarrow \Psi_{0,1}^{n+1}$, and, as
seen from Fig. 2f, this fact has a fundamental consequence: 
{\it at non-zero bias voltage the spin-current causes 
a net excess of population of one of the two degenerate noncollinear Neel states}. 
Thus, the neutral system is prepared in the $\left| -+\right\rangle$ state.

Finally, we consider a 3-center molecular wheel with local ZFS axes contained
in the molecular plane and tangential to the wheel's circumference. This system is
of special interest, being a model for the experimentally characterized 
lanthanide wheel Dy$_3$~\cite{dy3tang}, which has been recently shown to have almost tangential on-site 
anisotropy axes, leading to toroidal magnetization~\cite{dy3}. 
For simplicity, here we consider an analog of this system with $s=3/2$ on metal sites. The 
collective states are modelled by the noncollinear Ising Hamiltonian (\ref{hex}), with ferromagnetic 
exchange $J_I = 25$ cm$^{-1}$, and easy-axis ZFS parameter $D = 8 J_I$. 
The ground state of the 3-wheel is a doubly degenerate Kramer's doublet characterized by a toroidal magnetic moment
$\tau = \mu_B R \sum_p {\tilde s}_{z,p}$~\cite{prbrapid,dy3}, where $R$ is the radius of the triangle, and
$\mu_B$ is the Bohr magneton.  We denote the two states with $\left|+++\right\rangle$ 
($\tau= +9/2 R \mu_B$) and $\left|---\right\rangle$ ($\tau= -9/2 R \mu_B$), where the first position
refers to the atom more strongly bound to the ferromagnetic source, and the second position refers to the 
atom bound to the non-magnetic drain. 
The singly-charged system is investigated for $J_H=4 D$, and for
$t=0.05 D$ (adiabatic transfer) and  $t=D$ (strong-transfer).  The ground state of the singly charged 
system is always 4-fold degenerate.
The present spintronics setup (bottom of Fig. 1, spin-polarization axis of the source co-planar with the 
wheel's plane, and perpendicular to the bond between metal 1 and metal 2) implies that 
only those components of the charged ground state overlapping with the  
$\left|---\right\rangle$ toroidal state reported in Fig. 1 will be significantly populated,
by virtue of the Hund's coupling rule. In the adiabatic limit these states correspond to 
spin-switching states, {\it i.e.} to states which represent coherent hopping from center 1 to 
center 2, with inversion of spin-polarization, and found to be:
\begin{eqnarray}
\Psi_{0,1}^{n+1}  &\approx&  \left| --- \right\rangle 
\left( 
   a_1 \left|\up 0 0\right\rangle+ b_1 \left|0 \down 0\right\rangle
\right)\nonumber \\
\Psi_{0,2}^{n+1}  &\approx&  \left| --- \right\rangle \nonumber \\
  &\times&\left( a_2 \left|\up 0 0\right\rangle+ b_2 \left|0 \down 0\right\rangle 
+c_2 \left|0 0\up \right\rangle+ c_2 \left|0 0\down \right\rangle
\right)\nonumber 
\label{charge_toroidal}
\end{eqnarray}
with $\left| a_1\right|^2 = 0.4$, $\left| b_1\right|^2 = 0.38$, $\left| a_2\right|^2 = 0.21$,
$\left| b_2\right|^2 = 0.27$, and $\left| c_2\right|^2 = 0.19$.
In the non-adiabatic limit the weight of the spin-switching components becomes smaller (although does
not vanish), in favor of states representing spin-conserving hopping processes.

In figure 3a (adiabatic) and 3c (nonadiabatic) we report the spin current-voltage 
diagrams obtained for the two limits of the hopping parameter: as for the dimer system, we observe 
spin switching only in the adiabatic limit. However, due to the spin-polarization of the
source electrode, the population transfer from the $\left|---\right\rangle$ toroidal neutral state 
to the charged manifold always dominates the non-coherent kinetics, 
producing an excess of population of  $\left|+++\right\rangle$, in both weak and strong transfer limits 
(Fig. 3b and 3d). This demonstrates a viable spintronics strategy to prepare a non-zero density 
of toroidal molecular magnetization in the sample.

In conclusion, we have investigated spin-transport through noncollinear magnetic molecules in 
the sequential tunneling (CB) regime. Two fundamental phenomena are identified here. The first,
the {\it spin-switching effect}, is caused by the action of the noncollinear magnetization on 
the spin-current. The second, the {\it selective population bias} of one of the two partners of
a noncollinear doublet, is determined by the effect of a spin-current on the noncollinear states.
Non-collinearity is found to be the crucial ingredient in these phenomena. This work
represents the first step into the new domain of noncollinear molecular spintronics,
expected to have a significant impact on the quest for protected molecular qubits.

\end{document}